\documentclass[conference]{IEEEtran}
\IEEEoverridecommandlockouts
\usepackage{cite}
\usepackage{romannum}
\usepackage{amsmath,amssymb,amsfonts}
\usepackage{algorithm}
\usepackage{algpseudocode}
\usepackage{graphicx}

\usepackage{textcomp}
\usepackage{xcolor}
\def\BibTeX{{\rm B\kern-.05em{\sc i\kern-.025em b}\kern-.08em
    T\kern-.1667em\lower.7ex\hbox{E}\kern-.125emX}}
\begin{document}

\title{Modulation and signal class labelling using active learning and classification using machine learning}
\author{\IEEEauthorblockN{Bhargava B C}
\IEEEauthorblockA{\textit{Electronics and Communication} \\
\textit{NIT Surathkal}\\
Manglore, India \\
bhargavabc.203sp001@nitk.edu.in}
\and
\IEEEauthorblockN{Ankush Deshmukh}
\IEEEauthorblockA{\textit{Electronics and Communication} \\
\textit{NIT Surathkal}\\
Manglore, India \\
ankushsanjaydeshmukh.202sp003@nitk.edu.in}

\and

\IEEEauthorblockN{{A V Narasimhadhan}}
\IEEEauthorblockA{\textit{Electronics and Communication} \\
\textit{NIT Surathkal}\\
Manglore, India \\
dhan257@gmail.com}}

\maketitle

\begin{abstract}
Supervised learning in machine learning (ML) requires labelled data set. Further real-time data classification requires an easily available methodology for labelling. Wireless modulation and signal classification find their application in plenty of areas such as military, commercial and electronic reconnaissance and cognitive radio. This paper mainly aims to solve the problem of real-time wireless modulation and signal class labelling with an active learning framework. Further modulation and signal classification is performed with machine learning algorithms such as KNN, SVM, Naive bayes. Active learning helps in labelling the data points belonging to different classes with the least amount of data samples trained. An accuracy of 86 percent is obtained by the active learning algorithm for the signal with SNR 18 dB. Further, KNN based model for modulation and signal classification performs well over range of SNR, and an accuracy of 99.8 percent is obtained for 18 dB signal. The novelty of this work exists in applying active learning for wireless modulation and signal class labelling. Both modulation and signal classes are labelled at a given time with help of couplet formation from the data samples.

\end{abstract}


\section{Introduction}
A new era of research has emerged out of Artificial intelligence (AI). ML algorithms are seen as a subset of AI. ML algorithms are used to solve problems in image processing, speech processing, video processing, signal domain, wireless communication. Advancements of research in these fields have been happening for a decade now and further leading to a new era of learning and innovations. In this paper, an active learning based algorithm for real time class labelling with the help of buffer and selective sampling is proposed. Pre-trained Convolutional neural network (CNN) architectures are used for training purposes. The model is a sub-sampling based active learning framework for wireless signal class labelling. Further detailed methodology of proposed algorithm and reason behind the use of buffer and pre-trained model is explained in detail in following sessions.\\
Pre-existing well trained model, active learning with selective data points, and binary classifier model (one versus all). These stay in as basic three modules of this paper. CNN is the basic building block of pre-existing models used in this experimentation. CNN's have convolutional blocks that have weights in them, weights get updated during forward and backward propagation, these updates of weights help in learning the input data fed to the network. Learning in the forward direction is carried by weights and learning in the backward direction is carried with the help of gradients. CNN's are extensively used for training and getting results for image data sets and speech data frames too.\\

Selective sampling helps in picking up the most relevant data points from the data set. These most relevant data points in turn will help the model to label the rest of the data samples belonging to a particular class. Hence with the least amount of training data, samples can be classified accordingly. There have been many previous works on active learning in different applications such as Arabic text classification, biomedical image segmentation, etc. Usage of binary models in place of a single model for classification in this algorithm is due to benefits binary models or one versus all over other models. Binary models help in the addition of a new class in between the training or classification. Binary models are independent of each other and hence weight update, parametric functions such as cost function and balancing of a particular class does not affect other classes present in the data set.\\ 

The second part of this paper includes classification with KNN, Naive bayes, and SVM. The novelty of this work exists in the fact that both classes (modulation and signal) of an incoming signal are classified at a given time with the help of a simple KNN or ML algorithm. Further comparisons in terms of accuracy and space are made for ML-based algorithms for classification.\\
\section{Related works}
This session can be divided into two, firstly works carried out on active learning and second recent works carried on automatic modulation classification (AMC). Recent trends and works on active learning are explained in the following session.
Autoencoder's performance with and without active learning was experimented and results showed that active learning-enabled autoencoders were performing far better \cite{b1}. Autoencoders were used for the classification of hyperspectral images. Samples with the most relevance were selected through active learning. Autoencoders were trained with Kennedy space and India pines hyperspectral images, accuracy and maps for classification were derived out of model \cite{b1}.\\

Deep learning combined with active transfer learning was used for the classification purpose of hyperspectral images. Spectral features named deep joint spectral special features were extracted using stacked sparse autoencoders (SSAE) networks. The network was seen performing well with the least number of training samples due to active transfer learning. The proposed architecture was giving high accuracy percent with 20 percent of the data sample being trained. Hence active transfer learning-based network with SSAE was outperforming existing architectures for hyperspectral image classification \cite{b2}. Further, a detailed study on active learning and different strategies involved in active learning was made. Studies were done on methods of selection, strategies involved in the query, and applications of active learning. Various active learning methods such as batch model, multi-task active learning were discussed in detail \cite{b3}.\\

Deep active learning with pair-wise constraints (DPAC) was used to annotate and select the most significant samples for cancer cell detection. The most significant nucleus was selected out of data samples with the help of active learning. Pathology colon data set was used, With only about 60 percent of data samples being used, 79.2 percent was the F1 score obtained. The model was seen performing significantly well compared to preexisting models \cite{b4}. Detection of aircraft and segmentation problems has got its significance in military applications. Always a human expert cannot be made to sit to do the task of detection. Hence, a hybrid clustering-based active learning model was proposed to select data points that are most relevant during aircraft detection. The hybrid cluster model outperforms most of the existing active learning frameworks for aircraft detection problems \cite{b5}. Facial age estimation was performed with deep learning models with CNN and gradient descent methods. To improve the performance of the model active learning based feedback system was used. Relative feedback helped in identifying whether a given image was older or younger than the predicted output. The architecture was named as deep active learning with relative label feedback (DALRel) \cite{b6}. Test mean squared error function was seen to be settling faster compared to existing models.\\

The second part of the literature survey is automatic modulation classification. Automatic modulation classification with adversarial active transfer learning (ATLA) was carried and results were compared across models such as knn, decision tree, svm, and deep learning, further ATLA was seen performing better \cite{b7}. Constellation diagram based AMC was solved with the help of attentive siamese networks (ASN). Deep features were initially extracted with the help of CNN's sharing similar parameters. Performance analysis was made on the dataset with non-gaussian noise and an accuracy of 99 percent was obtained for SNR greater than 10 dB \cite{b8}. Cognitive radio finds its application in areas such as the internet of things. Automatic modulation classification in cognitive radio internet of things was carried out with the help of a stacked quasi recurrent neural network (S-QRNN). Results from S-QRNN were found with higher efficiency of 75.83 percent than the preexisting models and latency for execution being less than 59.31 percent \cite{b9}.\\

Modulation and signal classes of a given wireless signal were classified with the help of CNN and RNN-LSTM based model and further result in terms of accuracy was compared in detail with preexisting ML algorithms such as decision tree and random forest. CNN and RNN-LSTM based models were performing well with an accuracy of 99.36 percent. Further, a detailed parametric comparison was also done on available architectures in the letter \cite{b10}. Later two models were designed for modulation classification. The first one was the constellation image-based classification technique and the second one was classification with a graphic representation of features (GRF) \cite{b11}. GRF produces a spider graph for the features present in modulation schemes. GRF simulation results showed an accuracy of 86 percent for the signal with 10 dB SNR.\\

Later modulation classification was performed over the air condition with advancements in CNN layers. CNN with relu activation function and sigmoid activation function in the final layer was used to analyse fading channels such as rician, reyleigh \cite{b12}. Study regarding channel, information regarding time, and frequency of signal related to modulation is significant. Hence, a detailed study was made on best suiting frequency and time mechanism, channels that can give better performance with CNN architecture \cite{b13}. Labelling dataset remains a burden for supervised learning. Hence, semi supervised learning with encoder and CNN-LSTM combination network named semi AMC was developed. With the least amount of data samples trained, significant accuracy was obtained \cite{b14}. Semi AMC in particular was proposed for the modulation classification of radio signals.

\section{dataset and preprocessing}
Radcom dynamic is the data set used for modulation and signal class labelling. Further with the same dataset from 0 dB SNR to 18 dB SNR data samples are used for modulation and signal classification with ML algorithms. Modulation and signal class labels found together in the datasets are clubbed together to form doublets. Nine such doublets are formed. In an active learning algorithm, each of the data samples requires to be of dimension 224*224. Hence each data sample from 18 dB SNR is converted into a 2-dimensional image of 224*224. 
\section{Methodology}
Two methods are explained in this session. Initially, buffer enabled real-time active learning based neural network algorithm for class labelling. Second, ML-based algorithms such as KNN, Naive bayes, SVM for modulation and signal classification.
\subsection{Active learning proposed algorithm}

\begin{algorithm}
	\caption{Real time class labelling} 
	\begin{algorithmic}[1]
	\State {30 samples to be labelled by user.}
	\State {For every new class detected new binary model gets initialized.}
	\State Binary models are trained with one versus all method.
	\State Bulk edit stage.
	\State For each of 30 samples labelled by model, user reviews it and corrects the label if incorrect.
	\State Incorrect samples get accumulated in the buffer.
	\State In case buffer overflows then go to active learning to select the most relevant samples and thereby train the model with the most significant samples.
    \If{Predictions per each page is greater than 15\\} 
        iterate again from step 1 to step 7.
\Else  
        \State Go back to step number 5.
\EndIf 
        \State The whole data samples are collectively labelled by the user and algorithm.

	\end{algorithmic} 
\end{algorithm}
Initially, the dataset is fed to the model. 30 samples are to be labelled by a human oracle or the user. Next starts the training phase, for each of the new class labels detected, a binary model gets initialized. As discussed earlier pre-trained models are used for training. In this algorithm, MobileNet V-3 is the neural network architecture used for training and prediction. easily available weights are the main reason behind use of pre-trained neural network architectures. These initialized binary models are trained with one versus all methods to allow new classes coming in between classification. For every set of 30 samples labelled by the model, the user corrects the samples that are labelled incorrectly by the model. Further samples that are incorrectly labelled get accumulated in the buffer. In case buffer overflows that is several samples in the buffer exceed its capacity then go back to the active learning algorithm and train the model again with the most significant samples. In case predictions per page (predicted incorrectly) exceeds number 15 then start again and iterate over step 1 to step 7 of the algorithm. Else go back to step number 5 in the algorithm.
\subsection{ML algorithms for classification}
\subsubsection{KNN}
K- nearest neighbour algorithm is a machine learning algorithm for supervised learning-based multi class classification. KNN works based on a similarity between data points. The optimal value of k is found out and over the iteration similarity index between points is calculated and further points with most similarities are placed in a single class. KNN uses euclidean distance for the calculation of similarity between points. Suppose x and y are two data points with n dimension then the euclidean distance between them is calculated as follows, 
\begin{equation}
d(x,y)=\sqrt{(x_1-y_1)^2+........+(x_n-y_n)^2}\\
\label{eq1}
\end{equation}
\begin{equation}
d(x,y)=\sqrt{\sum_{i=1}^{n} (x_i-y_i)^2}\\
\label{eq2}
\end{equation}
\subsubsection{Naive Bayes}
Naive bayes is a supervised machine learning algorithm for multi-class classification works based on the bayes theorem. This classifier works purely based on a probabilistic model. Naive bayes assume each feature of a particular class to be independent of each other and classify points according to the bayes theorem. Posterior probabilities are calculated from the probability of likelihood and prior probability. There are different types of Naive bayes algorithms such as gaussian Naive bayes, bernoulli Naive bayes, categorical Naive bayes, etc. In this experiment gaussian Naive bayes is used for classification. Bayes theorem is given by the following equation,
\begin{equation}
p(\tfrac{x}{y})=\frac{p(y/x)*p(x)}{p(y)}\\
\label{eq3}
\end{equation}
$p(\tfrac{x}{y})$ is posterior probability, to be calculated.\\
p(y/x) is likelihood probability, already known from data.\\
p(x) is prior probability and p(y) is marginal probability.\\
\subsubsection{SVM}
Support vector machine is the most commonly used ML algorithm. This algorithm is used for classification and regression. SVM classifies given data points with help of hyperplanes. In a 2-dimensional space, hyperplane is a line that separates two sets of data points belonging to two classes. In higher dimensions, these lines that separate different classes are known as hyperplanes. SVM maximizes the margin (distance between closest point and hyperplane) and finds out the hyperplane. Data points closest to hyperplanes are known as support vectors. SVM with Gaussian kernel is used for classification. \\

All three of the above-mentioned ML algorithms are used for the modulation and signal classification of wireless signals. Simulations have been carried out in python and the scikit tool is used for simulating algorithms. Data points from 0 dB to 18 dB are being used for classification (belonging to nine different classes).

\section{Results}

\subsection{modulation and signal class labelling with active learning}
Fig. 1. shows the number of data samples labelled by the user and model. In Fig. 1. blue lines depict the number of samples labelled by the model and the orange line show amount of data samples labelled by the user. Out of the 5642 total number of 18-decibel data points, 4894 points are correctly labelled by model. Hence the model is giving an accuracy of 86.74 percent that is significant compared to the user labelling the whole data set. Further, Fig. 2. shows the performance of the proposed active learning algorithm over each iteration. As the number of iterations increases the model is labelling samples closer to 25 in a batch of 30. Fig. 3. explains the training time taken for the algorithm to run for each training number.
\begin{figure}[htbp]
\centerline{\includegraphics[scale=0.35]{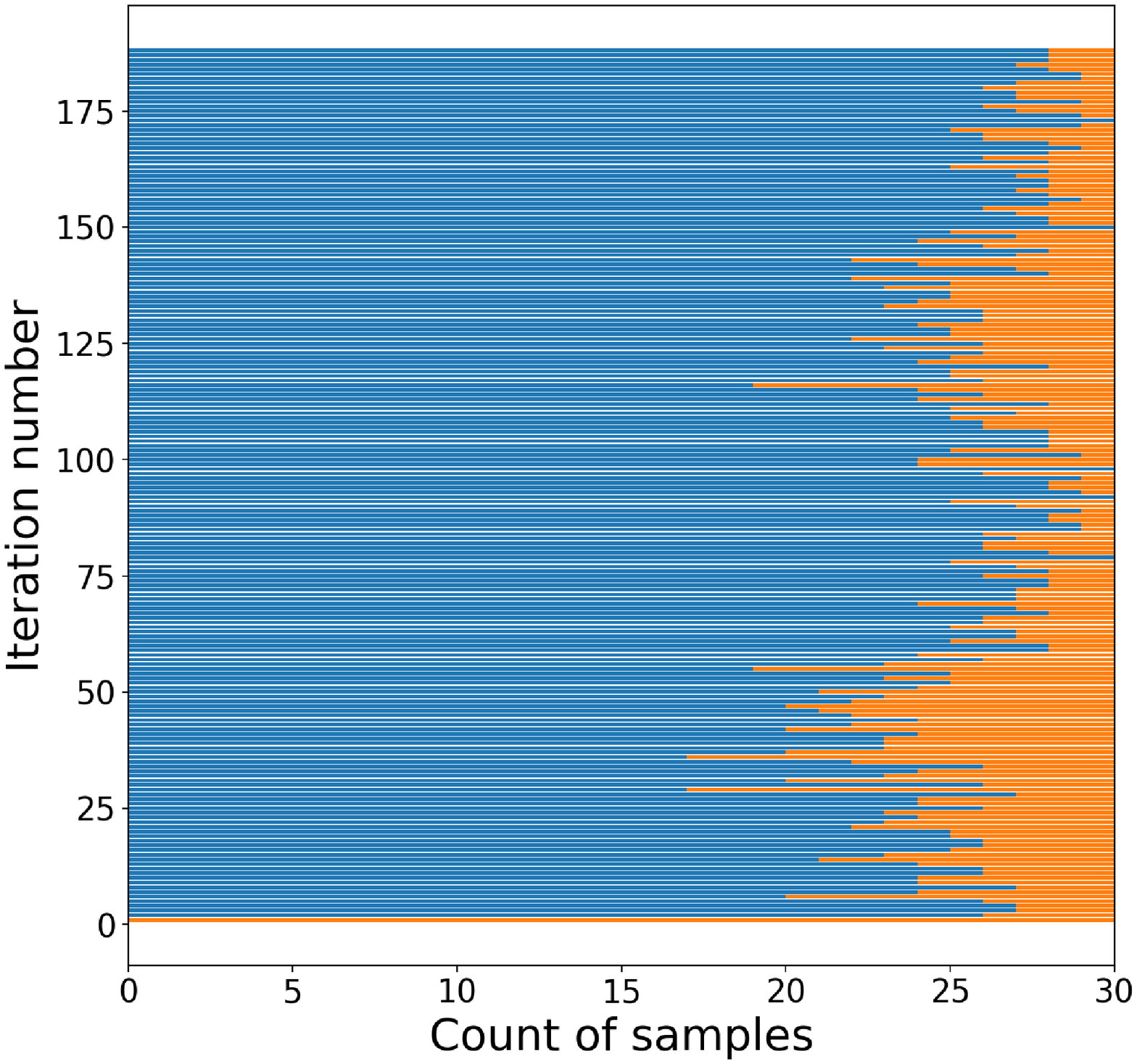}}
\caption{Model vs User labelling for 18 dB signal}
\label{fig .1}
\end{figure}
\begin{figure}[htbp]
\centerline{\includegraphics[scale=0.5]{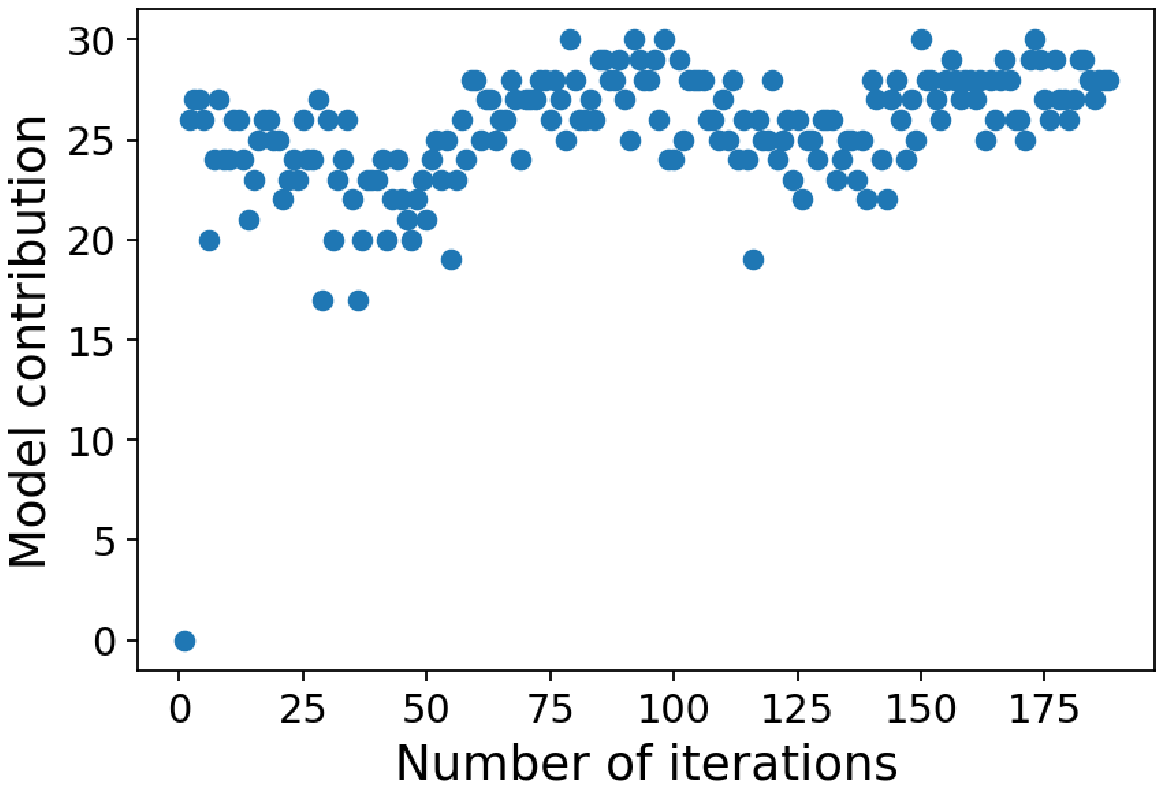}}
\caption{Model prediction for 18 dB}
\label{fig .2}
\end{figure}
\begin{figure}[htbp]
\centerline{\includegraphics[scale=0.5]{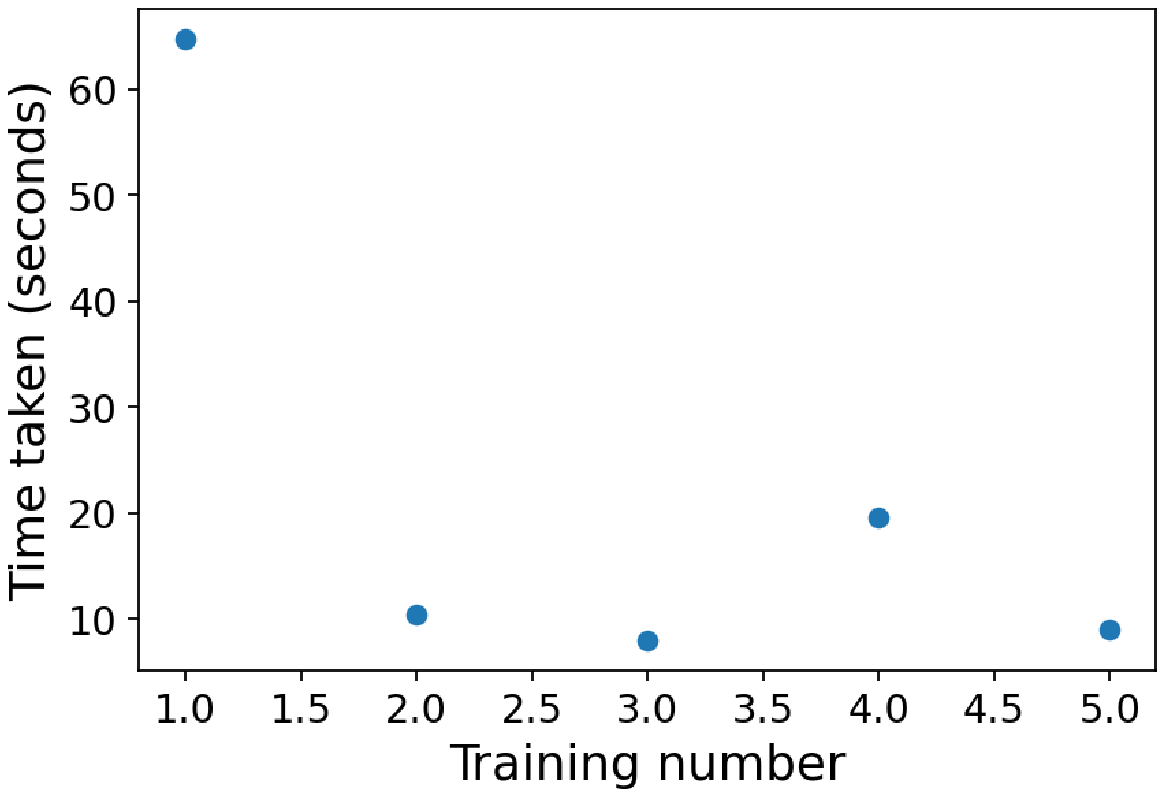}}
\caption{Model prediction for 18 dB}
\label{fig .3}
\end{figure}
\subsection{Classification with ML algorithms}
Fig. 4. shows the detailed comparison of results for wireless modulation and signal classification with ML algorithms such as K- nearest neighbor, Naive bayes, and Support vector machine. Graphical depiction does not give a picture of exact numerical values for each model. Table \Romannum{1} shows the detailed numerical values of the accuracy in percentage out of 100 for each of the above-mentioned ML algorithms used for modulation and signal classification. In terms of accuracy, KNN is performing better than Naive bayes and SVM models for classification in this case.
\begin{figure}[htbp]
\centerline{\includegraphics[scale=0.25]{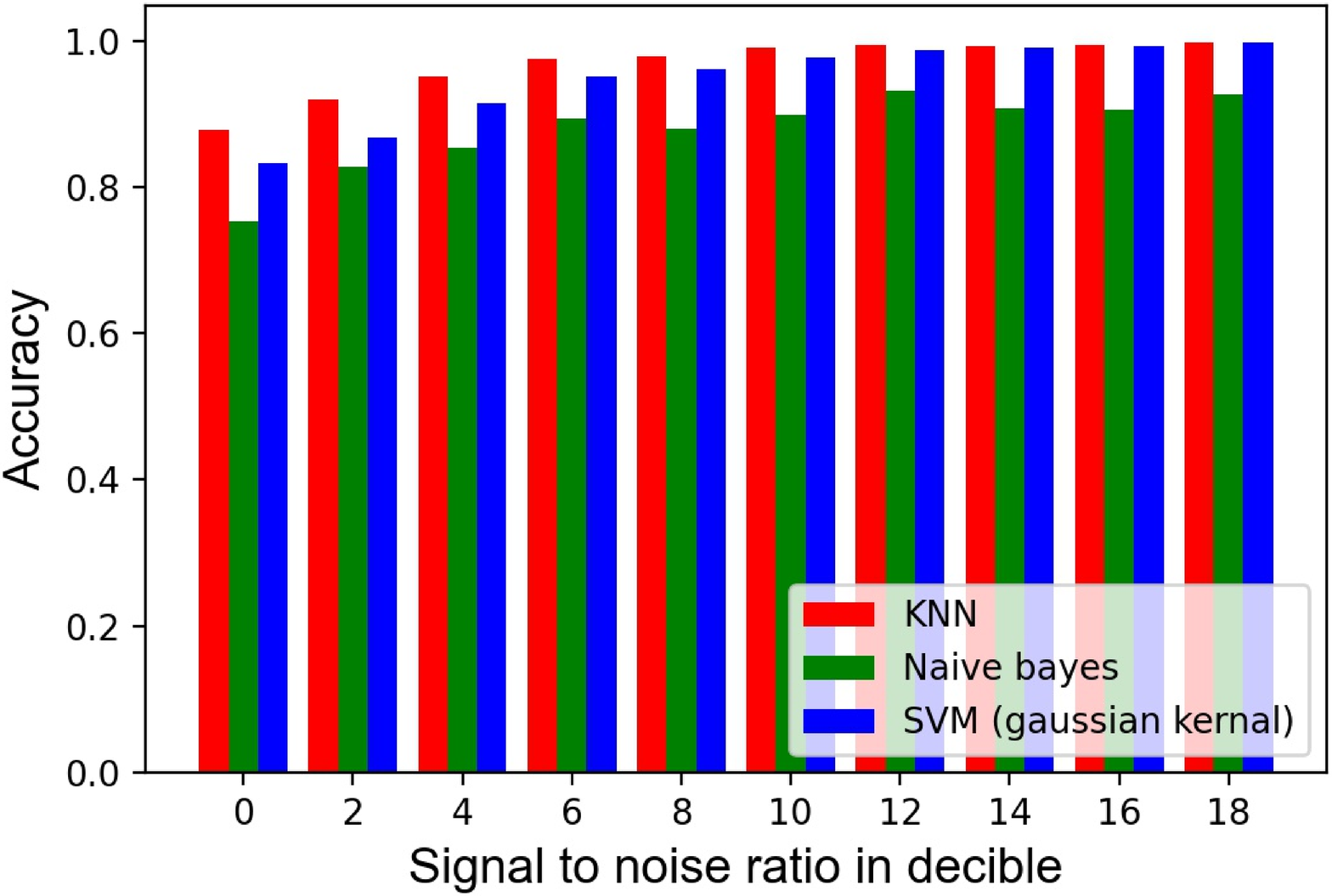}}
\caption{Model vs User labelling for 18 dB signal}
\label{fig .4}
\end{figure}
\begin{table}[ht]
\caption{Comparison of accuracy (in percentage) for different ML models}
\renewcommand{\arraystretch}{1.25}
\setlength{\tabcolsep}{15pt} 
\scriptsize 
\begin{tabular}{ |p{1cm}|p{1cm}|p{1cm}|p{1cm}|}
 \hline
\textbf{SNR (In decible)} & \textbf{KNN }& \textbf{Naive Bayes}  & \textbf{SVM}\\
\hline

 0 & \textbf{87.8 }& 75.3 & 83.3 \\
 \hline
 2 &  \textbf{91.9} & 82.8 &  86.7 \\
  \hline
  4 & \textbf{95.1} & 85.4  &  91.5 \\
  \hline
 6 & \textbf{97.5}  & 89.4 &  95.1 \\
  \hline
 8 & \textbf{97.9} & 87.9  & 96.1 \\
  \hline
 10 &\textbf{99.1}  & 89.9 & 97.6 \\
  \hline
 12 & \textbf{99.4} & 93.1  & 98.8 \\
 \hline
 14 &\textbf{99.3} & 90.8 & 99.1 \\
 \hline
 16 & \textbf{99.4} & 90.6 & 99.3 \\
 \hline
 18 & \textbf{99.8} & 92.7 & \textbf{99.8} \\
 \hline

\end{tabular}

\end{table}
\begin{figure}[htbp]
\centerline{\includegraphics[scale=0.25]{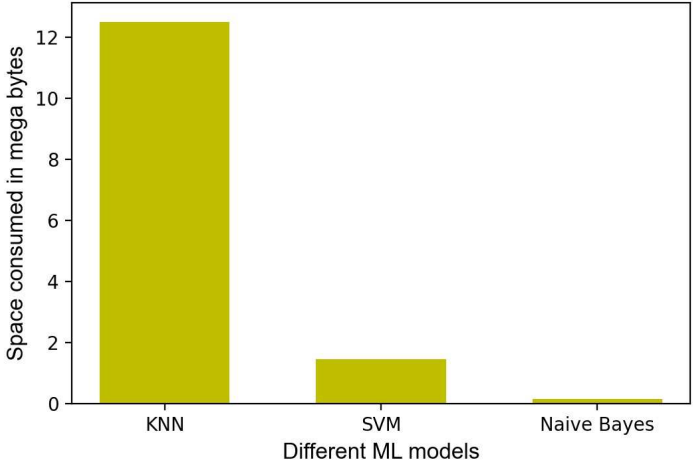}}
\caption{Space comparison for ML models}
\label{fig .5}
\end{figure}
Fig. 5. shows the comparison for KNN, Naive bayes, and SVM for space consumed in megabytes. KNN consumes 12.5 megabytes, SVM 1.49 megabytes and Naive bayes consume 50 kilobytes. Space consumed by each of the models is directly proportional to the number of parameters. Hence, KNN is the model with the highest number of parameters, and Naive bayes is the model with the least parameters.
\begin{table}[ht]
\caption{Comparison of KNN with state of art models }
\renewcommand{\arraystretch}{1.5}
\setlength{\tabcolsep}{10pt} 
\scriptsize 
\begin{tabular}{|p{1.8cm}|p{2cm}|p{2cm}|  }
 \hline
 \textbf{Model} & \textbf{Performance of the model} &\textbf{ Type of classification}  \\
 \hline
 KNN (proposed work) & $ 99.8\%$ for 18 dB SNR   & 6 Modulation 8 signal  \\
  \hline
 P. Ghasemzadeh, et al 2022 [9]&  75.83\% efficiency   & AMC in cognitive radio receiver  \\
  \hline
 Sun et al 2022 [11] & 86\% for 10 dB SNR & modulation classification with GRF\\
 \hline
  
\end{tabular}
\end{table}
Table \Romannum{2} compares KNN based model (proposed) to the previous works on modulation classification. The model performs better than state of art models for AMC.
\section{conclusion}
In this work, an attempt is made to solve the problem of real-time data labelling with active learning followed by classification of modulation and signal classes with KNN, Naive bayes, and SVM. Results from the active learning (proposed algorithm) show that the model can label 86.74 percent of the sample on its own. A novel attempt for a real-time buffer enabled algorithm for modulation and signal class labelling for 18 dB is giving significant accuracy. Active learning in wireless communication is a novel architecture that is proposed in this paper. Further moving forward, KNN is seen performing well for wireless modulation and signal classification, and an accuracy of 99.8 percent is obtained at 18 dB. KNN gives an accuracy of about 87.8 percent for 0 dB signal (a signal that has equal signal and noise strength). Although, SVM performs better than Naive bayes, not better than KNN. At 18 dB both KNN and SVM are giving an accuracy of 99.8 percent. Further KNN is performing very well with a wide range of SNR's ranging from 0 dB to 18 dB. \\
The future scope of this work is to build a model with an active learning-enabled classifier for real-time modulation and signal classification. Training and testing models with modulation classification of radio signals and cognitive radio is also a work that can be conducted in the future. The efficiency of the real-time active learning framework is to be checked on other neural network models such as Siamese network, GAN (generative adversarial network), CNN, LSTM (long short term memory network), etc.

\end{document}